\documentclass[aps]{revtex4}

\usepackage{graphicx}  % Include figure files
\usepackage{subfigure}
\usepackage{multirow}

\usepackage{fancyhdr}
\usepackage{longtable}
\usepackage{parskip}
\usepackage[T1]{fontenc}
\usepackage{dcolumn}   % Align table columns on decimal point

\usepackage{bm}        % bold math
\usepackage{amsfonts}  % Common math fonts
\usepackage{amsmath}   % Common math functions
\usepackage{amssymb}   % Common math symbols

\newcommand{\pwisein}{\left\{ \begin{array}{ll}}
\newcommand{\pwiseout}{\end{array}\right.}

\begin{document}

% --- TITLE ---
\title{Multi-Messenger Studies with High-Energy Neutrinos and Gamma Rays:\\ The WST Opportunity}

\author{Fabian Sch\"ussler$^{1}$, Sofia Bisero$^{1}$, Bernardo Cornejo$^{1}$, Filippo D'Ammando$^{2}$, Richard I. Anderson$^{3}$, Ilja Jaroschewski$^{1}$, Silvia Piranomonte$^{4}$, Fatemeh Zahra Majidi$^{5}$}

\affiliation{$^{1}$ \it IRFU, CEA, Université Paris-Saclay, F-91191 Gif-sur-Yvette, France}
\affiliation{$^{2}$ \it INAF-IRA Bologna, Italy}
\affiliation{$^{3}$ \it Institute of Physics, Ecole Polytechnique F\'ed\'erale de Lausanne (EPFL), Observatoire de Sauverny, 1290 Versoix, Switzerland}
\affiliation{$^{4}$ \it INAF-OAR, Italy}
\affiliation{$^{5}$ \it INAF OACN, Italy}

%\date{January 09, 2026}

% --- ABSTRACT / EXECUTIVE SUMMARY ---
\begin{abstract}
The search for the sources of ultra-high-energy cosmic rays (UHECRs) using high-energy neutrinos represents a frontier in high-energy astrophysics. However, a critical bottleneck remains: the ability to rapidly survey the sizable sky areas defined by the localization uncertainties of neutrino detectors and to provide rapid spectroscopic classification of the multitude of optical transients found within them. 

By deploying a large field-of-view with high-multiplex Multi-Object Spectroscopy (MOS) on a large aperture telescope, one can instantaneously cover neutrino error circles, thus providing crucial spectroscopic classifications of potential counterparts discovered, for example, by the Vera C. Rubin Observatory (LSST) with unprecedented efficiency. Furthermore, simultaneous operation of a giant panoramic central Integral Field Spectrograph (IFS) would allow for detailed kinematic and environmental characterization of primary candidates. This facility would unlock deep synergies between next-generation neutrino telescopes (IceCube-Gen2, KM3NeT) and gamma-ray observatories (CTAO), transforming unique multi-messenger alerts into a comprehensive physical understanding.
\end{abstract}

\maketitle

% --- MAIN BODY ---

\section{Scientific Context}
Ultra-high energy cosmic rays (UHECRs), particles with energies up to several $10^{20}$ eV, have puzzled the scientific community for over a century. While tremendous instrumental efforts have characterized average UHECR properties (energy spectrum, mass composition, anisotropies), the specific acceleration sites capable of reaching energies orders of magnitude beyond man-made accelerators remain elusive.

It has become evident that multi-messenger techniques are required to solve this puzzle. High-energy neutrinos, produced by hadronic interactions of UHECRs near their acceleration sites, travel unhindered through cosmic radiation fields, offering a direct probe of these violent phenomena. Yet, connecting a neutrino event to a specific astrophysical object is very challenging.

To date, only a handful of events have tentative multi-wavelength associations, such as the blazar TXS 0506+056, linked to the IceCube-170922A event \cite{IceCube2018a}. However, inconsistencies in modeling the full multi-messenger emission \cite{Keivani2018, Cerruti2019} and ``orphan'' neutrino flares \cite{IceCube2018b} highlight the complexity of these sources. Other neutrino events have been tentatively linked to Tidal Disruption Events (TDEs) \cite{Stein2021, Reusch2022}, necessitating a broader search capability.

The Vera C. Rubin Observatory (LSST) is starting to provide a wealth of faint transients within neutrino uncertainty regions \cite{Ivezic2019}. However, photometry alone is insufficient to establish the link with the neutrino emission. The bottleneck for the upcoming decade will be the \textbf{rapid spectral classification} of all transients within these regions to identify the true neutrino source. Assuming a one square-degree region defined by the neutrino localisation uncertainty, one can expect several hundreds of transients to be detected during a single snapshot by Rubin/LSST. While the majority of them can be identified as stellar flares and solar system objects, several tens of new transients will remain to be classified. Most of these will be supernovae, and the identification of their type and evolution stage is crucial information to eventually establish a statistical link with neutrino detections.

Similarly, in the Very High Energy (VHE) gamma-ray domain, instruments like Fermi-LAT and ground-based Cherenkov telescopes (H.E.S.S., MAGIC, VERITAS) are revealing transient TeV emission from Gamma-Ray Bursts \cite{HESS2019, MAGIC2019, HESS2021} and recurrent novae \cite{HESS2022}. While the angular resolution of the next-generation instrument CTAO will approach a few arcmin at the highest energies and for the strongest sources, faint or rapidly fading sources often lack the precise localization required for single-slit follow-up, and their host environments require detailed integral field analysis to understand the acceleration mechanisms at play.

\section{The Driving Questions}

\begin{enumerate}
    \item \textbf{What are the sources of UHECRs?} Exploiting the direct link between UHECR acceleration sites and emission of high-energy neutrinos: can we spectroscopically characterise the electromagnetic counterparts of high-energy neutrinos?
    \item \textbf{What physical processes accelerate particles to macroscopic energies?} How are multi-messenger signals (neutrinos, gravitational waves) and multi-wavelength emissions (from radio to TeV gamma-rays) physically linked in the environments of transient phenomena?
\end{enumerate}

\section{Technology Requirements}
To fully exploit the potential of the multi-messenger era, the next generation ESO observatory must operate in synergy with facilities like IceCube-Gen2, KM3NeT, and CTAO. This requires a specific set of technological capabilities to bridge the gap between detection and physical understanding. 

\begin{description}
    \item[Covering the Error Box (Large FoV + MOS)] Neutrino detectors typically produce error circles of the order of a square degree. To cover the entirety of these localization regions in a single pointing and eliminate the inefficiencies associated with tiling, a large Field of View (FoV) is essential. Furthermore, to simultaneously target and spectroscopically classify the numerous faint transients discovered by Rubin/LSST within these error boxes, a high-multiplex Multi-Object Spectrograph (MOS) provides the best tradeoff between coverage and FoV for poorly localized events, whereas a large (several square arcmin) IFS would be particularly powerful for better localized events or particularly crowded regions.
    \item[Matching the Survey Depth (Large Aperture)] To spectroscopically classify the transients detected, for example, by the Vera C. Rubin Observatory, which will typically be found at faint magnitudes ($r \sim 24$ mag), a large effective aperture is required. A massive collecting area is essential to achieve the necessary signal-to-noise ratio for these faint targets in reasonable exposure times, ensuring that the spectroscopic follow-up effectively matches the depth of the photometric discovery survey.
    \item[Characterizing the Physics (Medium \& High Res)] The next generation facility must offer both medium- and high-resolution modes. Specifically, a medium-resolution mode ($R \sim 4,000$) is essential for the rapid redshift determination and spectral typing of faint transients. Conversely, a high-resolution mode ($R \sim 40,000$) is required to resolve complex absorption profiles in the interstellar medium or map the detailed kinematics of bright counterparts. It is crucial to be able to rapidly obtain and then efficiently distribute this crucial information (hours to day timescales) to cover all potential transient phenomena and guide detailed follow-up observations throughout the electromagnetic spectrum.
    \item[Understanding the Environment (Simultaneous MOS and IFS)] Investigating the acceleration mechanisms of UHECRs requires more than just source identification; it demands a detailed characterization of the source's home environment. An IFS complements the broad search capability of a MOS by providing spatially resolved spectroscopy of the host galaxy. By obtaining this data alongside the MOS classification, observers can immediately link the transient event to the local gas dynamics and star formation history. This simultaneous, dual-mode approach offers the most efficient path to understanding the immediate environment of VHE neutrino and gamma-ray sources.
    \item[Future-Proofing] Upcoming neutrino telescopes will have higher sensitivities and thus reach higher redshifts. Still, we need to cover large (multiple square arcmin) regions even for the best-localized events. Maximizing spatial resolution for the IFS while maintaining a large FoV, e.g., via a ground-layer adaptive optics system, would allow us to follow up obscured sources and push to higher redshifts, matching improved multi-messenger capabilities.
\end{description}

We note that no existing facility currently offers this combination of large field-of-view, high-multiplex MOS, and simultaneous IFS on a large-aperture telescope. In this context, the conceptual design of the \textbf{Wide-field Spectroscopic Telescope} (WST, \cite{Mainieri2024}) is perfectly matched to these technological requirements.

\section*{Acknowledgements}
FS acknowledges the support of the French Agence Nationale de la Recherche (ANR) under reference ANR-22-CE31-0012 (project "Multi-messenger Observations of the Transient Sky (MOTS)").

% --- BIBLIOGRAPHY ---


\begin{thebibliography}{99}

\bibitem{IceCube2018a} IceCube Collaboration, ``Multimessenger observations of a flaring blazar coincident with high-energy neutrino IceCube-170922A'', \textit{Science} \textbf{361}, 1378 (2018).

\bibitem{Keivani2018} A. Keivani et al., ``A Multimessenger Picture of the Flaring Blazar TXS 0506+056'', \textit{ApJ} \textbf{864}, 84 (2018). 

\bibitem{Cerruti2019} M. Cerruti et al.,  ``Leptohadronic single-zone models for the electromagnetic and neutrino emission of TXS 0506+056'', \textit{MNRAS} \textbf{483}, 12 (2019).

\bibitem{IceCube2018b} IceCube Collaboration, ``Neutrino emission from the direction of the blazar TXS 0506+056 prior to the IceCube-170922A alert'', \textit{Science} \textbf{361}, 147 (2018).

\bibitem{Stein2021} R. Stein et al., ``A tidal disruption event coincident with a high-energy neutrino'', \textit{Nature Astronomy} \textbf{5}, 510 (2021). 

\bibitem{Reusch2022} S. Reusch et al., ``Candidate Tidal Disruption Event AT2019fdr Coincident with a High-Energy Neutrino'',\textit{PRL} \textbf{128}, 221101 (2022).

\bibitem{Ivezic2019} Z. Ivezic et al., ``LSST: From Science Drivers to Reference Design and Anticipated Data Products'', \textit{ApJ} \textbf{873} (2019).

\bibitem{HESS2019} H.E.S.S. Collaboration, ``A very-high-energy component deep in the $\gamma$-ray burst afterglow'', \textit{Nature} \textbf{575}, 464 (2019). 

\bibitem{MAGIC2019} MAGIC Collaboration,  ``Teraelectronvolt emission from the $\gamma$-ray burst GRB 190114C'', \textit{Nature} \textbf{575}, 455 (2019).

\bibitem{HESS2021} H.E.S.S. Collaboration, ``Revealing x-ray and gamma ray temporal and spectral similarities in the GRB 190829A afterglow'',  \textit{Science} \textbf{372}, 1081 (2021).

\bibitem{HESS2022} H.E.S.S. Collaboration, ``Time-resolved hadronic particle acceleration in the recurrent nova RS Ophiuchi'', \textit{Science} \textbf{376}, 77 (2022).

\bibitem{Mainieri2024} V. Mainieri et al., ``The Wide-field Spectroscopic Telescope (WST) Science White Paper'', arXiv:2403.05398 (2024).

\end{thebibliography}
\end{document}